\documentstyle[12pt]{article}
\def\nabstar#1{\nabla\kern-0.5pt\smash{\raise 4.5pt\hbox{$\ast$}}
               \kern-4.5pt_{#1}}

\def\drvstar#1{\partial\kern-0.5pt\smash{\raise 4.5pt\hbox{$\ast$}}
               \kern-5.0pt_{#1}}

\def\newline{\relax\ifhmode\null\hfil\break\else\nonhmodeerr@\newline\fi}
\def\frac#1#2{{#1\over#2}}
\def\text#1{{\hbox{\rm #1}}}

\newcommand{\beq}{\begin{equation}}
\newcommand{\eeq}{\end{equation}}
\newcommand{\bea}{\begin{eqnarray}}
\newcommand{\eea}{\end{eqnarray}}
\def\Id{ \mbox{1\hspace{-1.2mm}I} }
\def\BE{\begin{equation}}
\def\EE{\end{equation}}
\def\BA{\begin{eqnarray}}
\def\EA{\end{eqnarray}}
\def\BAN{\begin{eqnarray*}}
\def\EAN{\end{eqnarray*}}

\def\nn{\nonumber\\}

\def\tr{\mbox{tr}}
\def\Tr{\mbox{Tr}}

\def\gm5{\gamma_5}

\input epsf.sty
\newdimen\psfigsize
\def\psfigure#1 #2 #3 #4 #5{
    \begin{figure}[tbh]
      \begin{center}
      \vbox{
        \null\vskip-0.2in\hskip#2
        \epsfxsize=#1
        \epsfbox{#4}
        \vskip -0.3in
        \caption {#5 \label{#3}}
        \vskip 0.0 true in plus 0.3 true in
      }
      \end{center}
   \end{figure}
}
\begin{document}
\thispagestyle{empty}
\begin{flushright}
NTUTH-02-505A \\
April 2002
\end{flushright}
\bigskip\bigskip\bigskip
\vskip 2.5truecm
\begin{center}
{\LARGE {Quenched chiral logarithms in lattice QCD
         with exact chiral symmetry}}
\end{center}
\vskip 1.0truecm
\centerline{Ting-Wai Chiu and Tung-Han Hsieh}
\vskip5mm
\centerline{Department of Physics, National Taiwan University}
\centerline{Taipei, Taiwan 106, Taiwan.}
\centerline{\it E-mail : twchiu@phys.ntu.edu.tw}
\vskip 1cm
\bigskip \nopagebreak \begin{abstract}
\noindent

We examine quenched chiral logarithms in lattice QCD with overlap
Dirac quark. For 100 gauge configurations generated with the Wilson
gauge action at $ \beta = 5.8 $ on the $ 8^3 \times 24 $ lattice, we
compute quenched quark propagators for 12 bare quark masses.
The pion decay constant is extracted from the pion propagator, and
from which the lattice spacing is determined to be 0.147 fm.
The presence of quenched chiral logarithm in the pion mass is confirmed,
and its coefficient is determined to be $ \delta = 0.203 \pm 0.014 $,
in agreement with the theoretical estimate in quenched chiral perturbation
theory. Further, we obtain the topological susceptibility of these 100
gauge configurations by measuring the index of the overlap Dirac operator.
Using a formula due to exact chiral symmetry, we obtain the $ \eta' $ mass
in quenched chiral perturbation theory, $ m_{\eta'} = ( 901 \pm 64 ) $ MeV,
and an estimate of $ \delta = 0.197 \pm 0.027 $, which is in good agreement
with that determined from the pion mass.



\end{abstract}
\vskip 1.5cm

\newpage\setcounter{page}1

\section{Introduction}

In quenched chiral perturbation theory \cite{Sharpe:1992ft,Bernard:1992mk},
it has been asserted that some quantities in QCD possess
logarithmic dependence on the bare quark masses as
the latter approach zero, similar to the unquenched case.
In particular, for the pseudoscalar meson mass and
the condensate, there are extra chiral logarithms due to the
$ \eta' $ loop void of topological screening in the quenched approximation.
Explicitly, the pion mass to one-loop order can be written as
\bea
\label{eq:mpi2}
m_\pi^2 = C m_q \{ 1 - \delta[ \mbox{ln}( C m_q/\Lambda_{\chi}^2 ) + 1 ] \}
          + B m_q^2 \ ,
\eea
where $ m_q $ denotes the bare ( $ u $ and $ d $ ) quark mass,
$ \Lambda_{\chi} $ the chiral cutoff which can be taken to be
$ 2 \sqrt{2} \pi f_{\pi} $ ( $ f_{\pi} \simeq 132 $ MeV ),
$ C $ and $ B $ are parameters,
and $ \delta $ the coefficient of the quenched chiral logarithm.

Theoretically, $ \delta $ can be estimated to be \cite{Sharpe:1992ft}
\bea
\label{eq:delta}
\delta = \frac{ m_{\eta'}^2 }{ 8 \pi^2 f_\pi^2 N_f } \ ,
\eea
where $ m_{\eta'} $ denotes the $ \eta' $ mass in quenched
chiral perturbation theory, $ f_\pi $ the pion
decay constant, and $ N_f $ the number of light quark flavors.
For $ m_{\eta'} = 900 $ MeV, $ f_{\pi} = 132 $ MeV, and $ N_f = 3 $,
(\ref{eq:delta}) gives $ \delta \simeq 0.2 $. Evidently, if
one can extract $ \delta $ from the data of $ m_\pi^2 $ in lattice QCD,
then $ m_{\eta'} $ in quenched chiral perturbation theory
can be determined by (\ref{eq:delta}).

Presumably, the chiral logarithm in (\ref{eq:mpi2}) can be observed
unambiguously in quenched lattice QCD. However, it is rather difficult
to disentangle a logarithm from a power series.
So far, there is no compiling evidence to single out (\ref{eq:mpi2})
among many functional forms which seemingly can fit the data of
$ m_\pi^2 $ very well ( e.g., $ m_\pi^2 = A + C m_q + B m_q^2 $ ).

For lattice QCD with Wilson-Dirac quark, the chiral symmetry is explicitly
broken and the quark mass is additively renormalized, thus one does not
expect that quenched chiral logarithm can be identified unambiguously
at finite lattice spacing. In fact, if one tries to fit (\ref{eq:mpi2})
to the data of $ m_\pi^2 $ with Wilson-Dirac quark, one would obtain
the coefficient of quenched chiral logarithm,
$ \delta \simeq 0.06-0.12 $ \cite{Aoki:1999yr,Bardeen:2000cz},
quite smaller than its theoretical expectation $ 0.2 $. For lattice
QCD with staggered quark, the coefficient of chiral logarithm also
turns out to be rather small, $ \delta \simeq 0.06 $ \cite{Bernard:2001av},
similar to the case of Wilson-Dirac quark. These discrepancies indicate that
quenched chiral logarithm may not be properly reproduced
in lattice QCD with these two lattice fermion schemes,
at finite lattice spacing.

With the realization of exact chiral symmetry on the lattice, one
expects that quenched chiral logarithm can be reproduced in lattice
QCD with overlap Dirac quarks \cite{Neuberger:1998fp,Narayanan:1995gw}.
In a recent study \cite{Dong:2001fm},
the coefficient of quenched chiral logarithm
$ \delta $ is estimated to be in the range $ 0.15-0.4$ for
the chiral cutoff $ \Lambda_{\chi} $ in the range $ 0.6 - 1.0 $ GeV,
by fitting (\ref{eq:mpi2}) to their data of $ m_{\pi}^2 $.
At $ \Lambda_{\chi} a = 2 \sqrt{2} \pi f_{\pi} a = 1.17 $ GeV,
their data seems to give $ \delta \simeq 0.41(9) $, which is too large
comparing with the theoretical estimate, $ \delta \sim 0.18 - 0.20 $.
Thus, it is not clear whether quenched chiral logarithm has
been unambiguously identified.

Besides from the data of $ m_{\pi}^2 $,
one can also obtain $ \delta $ (\ref{eq:delta})
by extracting $ m_{\eta'} $ from the propagator
of the disconnected hairpin diagram.
However, to compute the propagator of hairpin is very tedious.
Fortunately, due to the chiral symmetry of $ D_c $
(\ref{eq:gen_sol}) in the quark propagator
$ ( D_c + m_q )^{-1} $, only the zero modes of $ D_c $ can contribute
to the hairpin diagram, regardless of the bare quark mass $ m_q $.
Thus one can derive an exact relation (\ref{eq:wv_lat}) between
the $ {\eta'} $ mass in quenched chiral perturbation
theory and the index susceptibility of any Ginsparg-Wilson lattice
Dirac operator, without computing the hairpin diagram explicitly.
Explicitly, it reads as
\bea
\label{eq:wv}
m_{\eta'}^2 = \frac{4 N_f }{f_{\pi}^2}
              \frac{\langle (n_{+}-n_{-})^2 \rangle }{V}
\eea
where $ N_f $ denotes the number of light quark flavors,
$ V $ the space-time volume, and
$ \langle (n_{+} - n_{+})^2 \rangle / V $ the index
susceptibility of any Ginsparg-Wilson lattice Dirac operator
in the quenched approximation.
Then (\ref{eq:delta}) and (\ref{eq:wv}) together gives
\bea
\label{eq:delta_s}
\delta
= \frac{ 1 }{ 2 {\pi}^2 ( f_{\pi} a )^4 }
  \frac{ \left< ( n_{+} - n_{-} )^2 \right> }{N_s}
\eea
where $ N_s $ is the total number of sites of the lattice.
A salient feature of (\ref{eq:delta_s}) is that $ \delta $
can be determined at finite lattice spacing $ a $, by measuring
the index ( susceptibility ) of the overlap Dirac operator, and
with $ f_{\pi} a $ extracted from the pion propagator.
Further, we suspect that the value of $ \delta $  (\ref{eq:delta_s})
is scale invariant at least for a range of lattice spacings
including the continuum limit $ a \to 0 $.

Now it is clear that, in order to confirm the presence of
quenched chiral logarithm in lattice QCD,
one needs to check whether the coefficient $ \delta $
determined from the data of $ m_{\pi}^2 $
agrees with that (\ref{eq:delta_s}) obtained from the index susceptibility.
Further, they should be close to the theoretical expectation
$ \delta \simeq 0.2 $. This is a requirement for the consistency of
the theory, since the quenched chiral logarithm in $ m_{\pi}^2 $
(\ref{eq:mpi2}) is due to the $ \eta' $ loop coupling to the
pion propagator through the mass term ( in the chiral lagrangian ),
thus $ \delta $ ( $ m_{\eta'} $ ) must be the same in both cases.
Otherwise, a consistent picture of the theory has not been established,
and one may infer that something must have gone wrong, either in the
formulation or in the implementation of the theory.
We regard this consistency requirement as a basic criterion for
lattice QCD ( with any fermion scheme ) to realize QCD chiral
dynamics in continuum.

In this paper, we examine whether lattice QCD with overlap Dirac quark
can satisfy this basic criterion.

\section{Pion mass and decay constant}

In this section, we set up our notations for computing quenched
quark propagator as well as meson propagators in lattice QCD
with exact chiral symmetry.
Then we describe the practical implementation for overlap
Dirac quark in our computations. Finally,
we present our results of $ m_\pi^2 a^2 $ and $ f_\pi a $ for 12
bare quark masses, and determine the coefficient of
quenched chiral logarithm, $ \delta = 0.203 \pm 0.014 $.

\subsection{Some general formulas}

First, we recall some general formulas in lattice QCD with exact
chiral symmetry. For any massless lattice Dirac operator $ D $
satisfying the Ginsparg-Wilson relation \cite{Ginsparg:1982bj}
\bea
\label{eq:gwr}
D \gamma_5 + \gamma_5 D = 2 r a D \gamma_5 D \ ,
\eea
it can be written in terms of a chirally symmetric Dirac operator
$ D_c $ \cite{Chiu:1998gp},
\bea
\label{eq:gen_sol}
D = D_c ( 1 + r a D_c )^{-1} \ .
\eea
Then the bare quark mass is naturally added to the $ D_c $
in the numerator of (\ref{eq:gen_sol}),
\bea
\label{eq:Dm}
D(m_q) = ( D_c + m_q )( 1 + r a D_c )^{-1} \ .
\eea
Thus the quark propagator is
\bea
\label{eq:Dcm}
( D_c + m_q )^{-1} = ( 1 - r m_q a )^{-1} [ D^{-1}(m_q) - r a ] \ ,
\eea
which is chirally symmetric in the massless limit. This exact chiral
symmetry is the crucial feature one would like to preserve for any quark
coupling to physical hadrons \cite{Chiu:1998eu}.

Then the pion propagator can be written as
\bea
\label{eq:pion}
M(\vec{x},t;\vec{0},0) =
\tr \{ \gamma_5 ( D_c + m_q )^{-1} ( 0, x ) \gamma_5
                ( D_c + m_q )^{-1} ( x, 0 ) \} \ ,
\eea
where the trace $ \tr $ runs over Dirac and color space.
Note that any one of the quark propagators in (\ref{eq:pion}) can
form the $ \eta' $ loop which leads to
the quenched chiral logarithm in the pion mass as well as the condensate.

With the pion propagators, one can compute its time
correlation function
\bea
\label{eq:Gt}
G(t) = \sum_{ \vec{x} } M(\vec{x},t;\vec{0},0) \ ,
\eea
which can then be fitted by the usual formula
\bea
\label{eq:Gt_fit}
G_{\pi}(t) = \frac{Z}{2 m_{\pi} a }
             [ e^{-m_{\pi} a t} + e^{-m_{\pi} a (T-t)} ] \ ,
\eea
to extract pion mass $ m_{\pi} a $,
and pion decay constant,
\bea
\label{eq:fpi}
f_{\pi} a = 2 m_q a \frac{\sqrt{Z}}{m_{\pi}^2 a^2 } \ .
\eea

Besides the exact chiral symmetry, we also require that
$ i D_c $ is hermitian, similar to the Dirac operator
$ i \gamma_\mu ( \partial_\mu + i A_\mu ) $ in continuum.
Then we have $ D_c^{\dagger} = \gamma_5 D_c \gamma_5 $.
As a consequence of hermiticity and chiral symmetry, both quark
propagators in (\ref{eq:pion}) can be expressed in terms of that
propagating from $ ( \vec{0}, 0 ) $ to $ ( \vec{x}, t ) $, i.e.,
\bea
\hspace{-4mm} M(\vec{x},t;\vec{0},0)
&=&
 \tr \{ [( D_c + m_q )^{-1} ( 0, x ) ]^{\dagger}
              ( D_c + m_q )^{-1} ( x, 0 ) \}  \nn
\hspace{-4mm} &=& \sum_{a,b=1}^3 \sum_{\alpha,\beta=1}^4
 \left[{( D_c + m_q )^{-1}}_{a\alpha}^{b\beta} ( x, 0 ) \right]^{*}
 {( D_c + m_q )^{-1}}_{a\alpha}^{b\beta} ( x, 0 )
\label{eq:pion_1}
\eea
where $ a, b $ denote color indices, and $ \alpha, \beta $ Dirac indices.
In other words, to obtain the time correlation function
(\ref{eq:Gt}) of pion, one only needs to compute 12 columns
of the matrix $ ( D_c + m_q )^{-1} $.

\subsection{Practical implementation for overlap Dirac quark}

The massless overlap Dirac operator \cite{Neuberger:1998fp} reads as
\bea
\label{eq:overlap}
D = m_0 a^{-1} \left( \Id + \gamma_5 \frac{H_w}{\sqrt{H_w^2}} \right)
\eea
where $ H_w $ denotes the hermitian Wilson-Dirac operator with a
negative parameter $ -m_0 $,
\bea
\label{eq:Hw}
H_w = \gamma_5 D_w = \gamma_5 (-m_0 + \gamma_\mu t_\mu + W ) \ ,
\eea
$ \gamma_\mu t_\mu $ the naive fermion operator, and $ W $
the Wilson term. Then $ D $ (\ref{eq:overlap}) satisfies the
Ginsparg-Wilson relation (\ref{eq:gwr}) with $ r = 1/(2 m_0 ) $.
Note that the value of $ m_0 $ has to be chosen properly such that the
overlap Dirac operator can capture the topology of the gauge configuration,
as well as behaving like a massless Dirac fermion even in a topologically
trivial gauge background. In other words, on a finite lattice, the
proper range of $ m_0 $ for any gauge configuration is smaller than that
$ ( 0 < m_0 < 2 ) $ in the free fermion limit. However, one usually does
not need to finely tune $ m_0 $ except for rough gauge configurations at
strong couplings \cite{Chiu:1998ce}.
In this paper, we fix $ m_0 = 1.3 $ for our computations.

Now the difficult task is to compute the quark propagator
with exact chiral symmetry (\ref{eq:Dcm}), which amounts to computing
\BAN
D^{-1}(m_q)= D^{\dagger}(m_q) \{ D(m_q) D^{\dagger}(m_q) \}^{-1} \ ,
\EAN
where the positive hermitian operator $ D(m_q) D^{\dagger}(m_q) $
can be written as
\bea
\label{eq:H2}
D(m_q) D^{\dagger}(m_q) =
\left[ m_q^2 + \left( m_0^2 - \frac{m_q^2}{4} \right)
               \left( 2 + \gamma_5 \frac{H_w}{\sqrt{H_w^2}}
             + \frac{H_w}{\sqrt{H_w^2}} \gamma_5 \right) \right] \ .
\eea
Then the quark propagators from the reference point $ ( \vec{0}, 0 ) $ to
all sites $ ( \vec{x}, t ) $ can be obtained by solving 12
column vectors $ Y_{c\alpha}(\vec{x},t) $ ( where $ c $ and
$ \alpha $ are color and Dirac indices of the quark field
at $ (\vec{0},0) $ ) in the following linear system
for 12 point source vectors on the r.h.s.,
\bea
\label{eq:H2Y}
 D(m_q) D^{\dagger}(m_q) Y_{c\alpha}(\vec{x},t) = \Id_{c\alpha} \ ,
\eea
where $ D(m_q) D^{\dagger}(m_q) $ is expressed by (\ref{eq:H2}).
Due to the rather small size of the physical memory in the present
generation of computers, the viable ways to solve the linear
system (\ref{eq:H2Y}) are iterative methods, in which the conjugate
gradient algorithm is the most optimal for positive definite matrices.
Starting from an inital guess of $ Y^{(0)} $, we multiply
$ D D^{\dagger} $ to $ Y^{(0)} $,
and update $ Y^{(0)} $ to $ Y^{(1)} $ according to the conjugate
gradient algorithm, then iterate this process until $ Y^{(n)} $
converges to the solution with desired accuracy $ \epsilon $, i.e.,
\bea
\label{eq:epsilon}
|| D(m_q) D^{\dagger}(m_q) Y^{(n)}_{c\alpha}(\vec{x},t) - \Id_{c\alpha} ||
< \epsilon \ .
\eea
In this paper, we set $ \epsilon = 10^{-11} $.

Now the updating process involves the multiplication of the matrix
\bea
\label{eq:inv_sqrt}
\frac{1}{\sqrt{H_w^2}}
\eea
to the vector $ Y^{(i)} $. However, (\ref{eq:inv_sqrt})
does not have analytic closed form.
This is the major challenge for lattice QCD with overlap Dirac quark.
A way to proceed is to express (\ref{eq:inv_sqrt}) in
terms of a rational approximation \cite{Neuberger:1998my}
\bea
\label{eq:rational}
\frac{1}{\sqrt{H_w^2}} \simeq \sum_{l=1}^k \frac{b_l}{H_w^2 + d_l} \ ,
\eea
where $ b_l $ and $ d_l $ are some positive definite coefficients.
Then the matrix-vector multiplication
\BAN
\frac{1}{\sqrt{H_w^2}} Y^{(i)} \simeq
\sum_{l=1}^k \frac{b_l}{H_w^2 + d_l} Y^{(i)} = \sum_{l=1}^k b_l Z_l
\EAN
can be evaluated by invoking another conjugate gradient process
to the linear systems
\bea
\label{eq:inner}
( H_w^2 + d_l ) Z_{l} = Y^{(i)}, \hspace{4mm} l = 1, \cdots, k.
\eea

Instead of solving each $ Z_l $ individually, one can use
multi-shift CG algorithm \cite{Frommer:1995ik}, and obtain all $ Z_l $
altogether, with only a small fraction of the total time what one had
computed each $ Z_l $ separately. Evidently, one can also apply
multi-shift CG algorithm to (\ref{eq:H2Y}) to obtain several quark
propagators with different bare quark masses.

Now the computation of overlap Dirac quark propagator involves
two nested conjugate gradient loops : the so-called inner CG loop
(\ref{eq:inner}), and the outer CG loop (\ref{eq:H2Y}).
The inner CG loop is the price what one pays for preserving the
exact chiral symmetry at finite lattice spacing.

The computational cost can be saved more than $ 50\% $ by
observing that $ D(m_q) D^{\dagger}(m_q) $ (\ref{eq:H2}) commutes with
$ \gamma_5 $. Thus one can choose chiral sources on the r.h.s.
of (\ref{eq:H2Y}),
\BAN
(\Id_{c\alpha} )^{b\beta}(\vec{x},t)
= \delta_{cb} \delta_{\alpha\beta} \ p( \vec{x}, t )
\EAN
where $ p(\vec{x},t) = \delta_{\vec{x},\vec{0}} \delta_{t,0} $
for a point source. Then we have
\BAN
\gamma_5 \Id_{c\alpha} &=& + \Id_{c\alpha}, \hspace{4mm} \alpha = 1,2 \ , \\
\gamma_5 \Id_{c\alpha} &=& - \Id_{c\alpha}, \hspace{4mm} \alpha = 3,4 \ ,
\EAN
and
\BAN
\gamma_5 Y_{c\alpha} &=& + Y_{c\alpha}, \hspace{4mm} \alpha = 1,2 \ , \\
\gamma_5 Y_{c\alpha} &=& - Y_{c\alpha}, \hspace{4mm} \alpha = 3,4 \ .
\EAN
Now one of the two matrix-vector multiplications in the outer CG loop
can be eliminated, since the l.h.s. of (\ref{eq:H2Y}) becomes
\bea
\label{eq:H2YC}
D(m_q) D^{\dagger}(m_q) Y_{c\alpha} =
\left\{  m_q^2 + \left( 2 m_0^2 - \frac{m_q^2}{2} \right)
\left[1+\frac{(\gamma_5 \pm 1)}{2} H_w \frac{1}{\sqrt{H_w^2}} \right] \right\}
Y_{c\alpha}
\eea
where `+' for $ \alpha = 1,2 $, and `-' for $ \alpha = 3,4 $.
Further, due to the projector
$ ( \gamma_5 \pm 1 )/2 $, half of the matrix-vector multiplication in
\BAN
 H_w \cdot \frac{1}{\sqrt{H_w^2}} Y_{c\alpha}
\EAN
can be saved.

The next crucial question is how to fix the values of coefficients
$ b_l $ and $ d_l $ in (\ref{eq:rational}) such that the inverse square
root function can be approximated optimally.
Fortunately, this problem has been solved by Zolotarev \cite{Zol:1877}
in 1877, using Jacobian elliptic functions.
A comparative study of Zolotarev's optimal rational approximation versus
other schemes for the overlap Dirac operator has been reported in
Ref. \cite{vandenEshof:2001hp}.

For the inverse square root function
\bea
\frac{1}{\sqrt{x}}, \hspace{4mm}  x \in [ 1, x_{max} ] \ ,
\eea
the Zolotarev optimal rational approximation is
\bea
\label{eq:fz}
f(x) = d_0 \frac{\prod_{l=1}^{k-1} (x+c_{2l})}{\prod_{l=1}^k ( x + c_{2l-1} )}
\eea
where
\bea
c_l = \frac{\mbox{sn}^2( lK/2k; \kappa ) }{1-\mbox{sn}^2( lK/2k; \kappa)},
\hspace{4mm} \kappa = \sqrt{ 1 - 1/x_{max} } \ ,
\eea
the Jacobian elliptic function $ \mbox{sn}(u,\kappa) = \eta $ is defined by
the elliptic integral
\bea
u(\eta) = \int_{0}^{\eta} \frac{dt}{\sqrt{(1-t^2)(1- \kappa^2 t^2 )}} \ ,
\eea
$ K=u(1) $ is the complete elliptic integral, and
$ d_0 $ is uniquely determined by the condition
\BAN
    \mbox{max} [ 1 - \sqrt{x} f(x) ] |_{ x \in [1,x_{max}] }
= - \mbox{min} [ 1 - \sqrt{x} f(x) ] |_{ x \in [1,x_{max}] }
\EAN
For our purpose, it suffices to fix $ d_0 $ such that $ f(1)=1 $.

Since the eigenvalues of $ H_w $ are bounded,
\bea
\label{eq:bounds}
\lambda_{min} \le | \lambda(H_w) | \le \lambda_{max} \ ,
\eea
we can rescale $ H_w $,
\bea
\label{eq:hw}
h_w \equiv H_w / \lambda_{min} \ ,
\eea
such that the eigenvalues of $ h_w^2 $ fall inside the
interval $ [ 1, (\lambda_{max}/\lambda_{min})^2 ] $.

After obtaining the partial fraction of (\ref{eq:fz}), we have
the optimal rational approximation to the inverse square root
of $ H_w^2 $
\bea
\label{eq:inv_Hw2}
\frac{1}{\sqrt{H_w^2}} \simeq \frac{1}{\lambda_{min}} \
                              \sum_{l=1}^k \frac{b_l}{ h_w^2 + d_l }
\eea
where
\bea
d_l &=& c_{2l-1} \ , \\
b_l &=& d_0 \frac{ \prod_{i=1}^{k-1} ( c_{2i} - c_{2l-1} ) }
             { \prod_{i=1, i \ne l}^{k} ( c_{2i-1} - c_{2l-1} ) } \ ,
\eea
and the parameter $ \kappa $ in the Jacobian elliptic function
$ \mbox{sn}( lK/2k; \kappa ) $ is
\bea
\label{eq:kappa}
\kappa = \sqrt{ 1 - (\lambda_{min}/\lambda_{max})^2 } \ .
\eea

Note that, for a given order $ k $, the smaller
the interval $ [ 1, (\lambda_{max}/\lambda_{min})^2 ] $,
the more accurate the rational approximation is.
Thus, it is advantageous to narrow the interval
$ [ \lambda_{min}, \lambda_{max} ] $ by projecting out the
eigenmodes at both ends. Since the spectrum of $ H_w^2 $ is very dense
near the upper bound, we only need to project out one or two
eigenmodes near $ \lambda_{max} $, just to get the precise value
of $ \lambda_{max} $.
On the other hand, we usually project out 20 or more low-lying eigenmodes
of $ H_w^2 $, such that $ (\lambda_{max}/\lambda_{min})^2 $ is always
less than 2500. Then we project the inner CG loop to the complement of
the vector space spanned by these eigenmodes.
This not only gives a better rational approximation for the inverse square
root of $ H_w^2 $, but also reduces the number of iterations of the inner
CG loop, which is proportional to the condition number
$ (\lambda_{max}/\lambda_{min})^2 $.

We use Arnoldi algorithm to compute a selected subset of eigenmodes of
$ H_w $, which correspond to high and low lying ones of $ H_w^2 $.
Denoting these eigenmodes by
\bea
\label{eq:eigen}
H_w u_j = \lambda_j u_j, \hspace{4mm} j =1, \cdots, n \ ,
\eea
we project the inner CG loop (\ref{eq:inner}) to the complement of the vector
space spanned by these eigenmodes
( i.e., multiplying both sides of (\ref{eq:inner}) by the projector
$ P = 1 - \sum_{j=1}^n u_j u_j^{\dagger} $, and use $ P H_w = H_w P $ ),
\bea
\label{eq:inner_p}
( h_w^2 + d_l ) \bar{Z}_{l} = \bar{Y}^{(i)}
\equiv ( 1 - \sum_{j=1}^n u_j u_j^{\dagger} ) Y^{(i)} \ .
\eea
Note that $ \lambda_{min} $ and $ \lambda_{max} $ in
Eqs. (\ref{eq:bounds})-(\ref{eq:kappa})
now refer to upper and lower bounds of $ | \lambda(\bar{H}_w) | $,
where
\bea
\label{eq:Hbar}
\bar{H}_w = P H_w = H_w - \sum_{j=1}^n \lambda_j u_j u_j^{\dagger} \ .
\eea

Then the matrix-vector multiplication at each step of outer CG loop
can be expressed in terms of the projected eigenmodes (\ref{eq:eigen})
plus the solution obtained from the inner CG loop (\ref{eq:inner_p})
in the complementary vector space, i.e.,
\bea
\label{eq:epsilon_Y}
H_w \frac{1}{\sqrt{H_w^2}} Y^{(i)} \simeq
  \frac{1}{\lambda_{min}} \bar{H}_w \sum_{l=1}^k b_l \bar{Z}_l +
\sum_{j=1}^n \frac{\lambda_j}{\sqrt{\lambda_j^2}} u_j u_j^{\dagger} Y^{(i)}
\equiv S^{(i)} \ .
\eea
The accuracy of $ S^{(i)} $ can be measured by the deviation
\bea
\label{eq:sign_acc}
\sigma_i = \frac{|{S^{(i)}}^{\dagger} S^{(i)} - {Y^{(i)}}^{\dagger} Y^{(i)}| }
{ {Y^{(i)}}^{\dagger} Y^{(i)}} \ ,
\eea
which would be zero if (\ref{eq:epsilon_Y}) is exact.

In this paper, we project out 20 lowest-lying and 4 largest
eigenmodes of $ H_w^2 $, and use 20 terms ( $ k=20 $ ) in the
Zolotarev optimal rational approximation, and set the convergence
criterion for inner CG loop to $ 10^{-12} $,
then $ \sigma_i $ is always less than
$ 10^{-11} $ for any iteration of the outer CG loop.

\subsection{Results}

With the Wilson $ SU(3) $ gauge action at $ \beta = 5.8 $, we generate 100
gauge configurations on the $ 8^3 \times 24 $ lattice, using
Creutz-Cabibbo-Marinari heat bath algorithm \cite{Creutz:zw,Cabibbo:zn}.
For each configuration, we compute quenched quark propagators for
12 bare quark masses. Then the pion propagator (\ref{eq:pion_1}) and its
time correlation functions (\ref{eq:Gt}) are obtained, and the latter are
fitted by (\ref{eq:Gt_fit}) to yield the pion mass $ m_\pi a $ and
decay constant $ f_\pi a $.

In Figs. \ref{fig:mpi2} and \ref{fig:fpi}, we plot
the pion mass square $ ( m_{\pi} a )^2 $ and decay constant
$ f_{\pi} a $ versus the bare quark mass $ m_q a $ respectively.
The data of $ f_{\pi} a $ can be fitted by
\bea
\label{eq:fpi_fit}
f_{\pi} a = 0.0984(3) + 0.1635(15)  ( m_q a )
\eea
with $ \chi^2 / d.o.f. = 0.03 $. Now we take the value
$ f_{\pi} a = 0.0984(3) $ at $ m_q a = 0 $ equal to the experimental value
of pion decay constant $ f_\pi = 132 $ MeV times the lattice spacing $ a $,
\bea
\label{eq:fpi_a}
f_{\pi} a = 0.0984(3) = 132 \mbox{ MeV } \times a \ ,
\eea
then this gives an estimate of the lattice spacing
\bea
\label{eq:a}
a = 0.147(1) \mbox{ fm }.
\eea
Thus the lattice size is $ \sim ( 1.2 \mbox{ fm } )^3 \times 3.5 \mbox{ fm } $.
Since the smallest pion mass is $ \sim 418 $ MeV,
the lattice size is
$ \sim (2.5)^3 \times 7.5 $, in units of the
Compton wavelength ( $ \sim 0.47 $ fm ) of the lowest-mass pion.

Fixing the cutoff $ \Lambda_{\chi} =  2 \sqrt{2} \pi f_\pi $
as in quenched chiral perturbation theory
( i.e., $ \Lambda_{\chi} a = 2 \sqrt{2} \pi f_\pi a = 0.8745 $ ),
we fit (\ref{eq:mpi2}) to our data of $ ( m_\pi a )^2 $, and obtain
\bea
\label{eq:delta_pi}
\delta &=& 0.2034(140) \ , \\
C a &=& 1.1932(182) \ ,  \\
\label{eq:B}
B &=& 1.1518(556)  \ ,
\eea
with $ \chi^2 / d.o.f. = 0.03 $.

Even though one may not easily detect the quenched chiral logarithm in
Fig. \ref{fig:mpi2}, it can be unveiled by
plotting $ ( m_\pi a )^2 / ( m_q a ) $ versus $ m_q a $, as shown in
Fig. \ref{fig:mpi2omq}. Further, we can subtract the quadratic term
$ B m_q^2 $ [ with the value of $ B $ given in (\ref{eq:B}) ],
and check its dependence on $ \ln (m_q) $ explicitly.
In Fig \ref{fig:chilog}, we plot
$ ( m_\pi a )^2 / ( m_q a ) - B ( m_q a ) $
versus $ \log(m_q a) $. The presence of quenched chiral
logarithm is evident.

\psfigure 5.0in -0.2in {fig:mpi2} {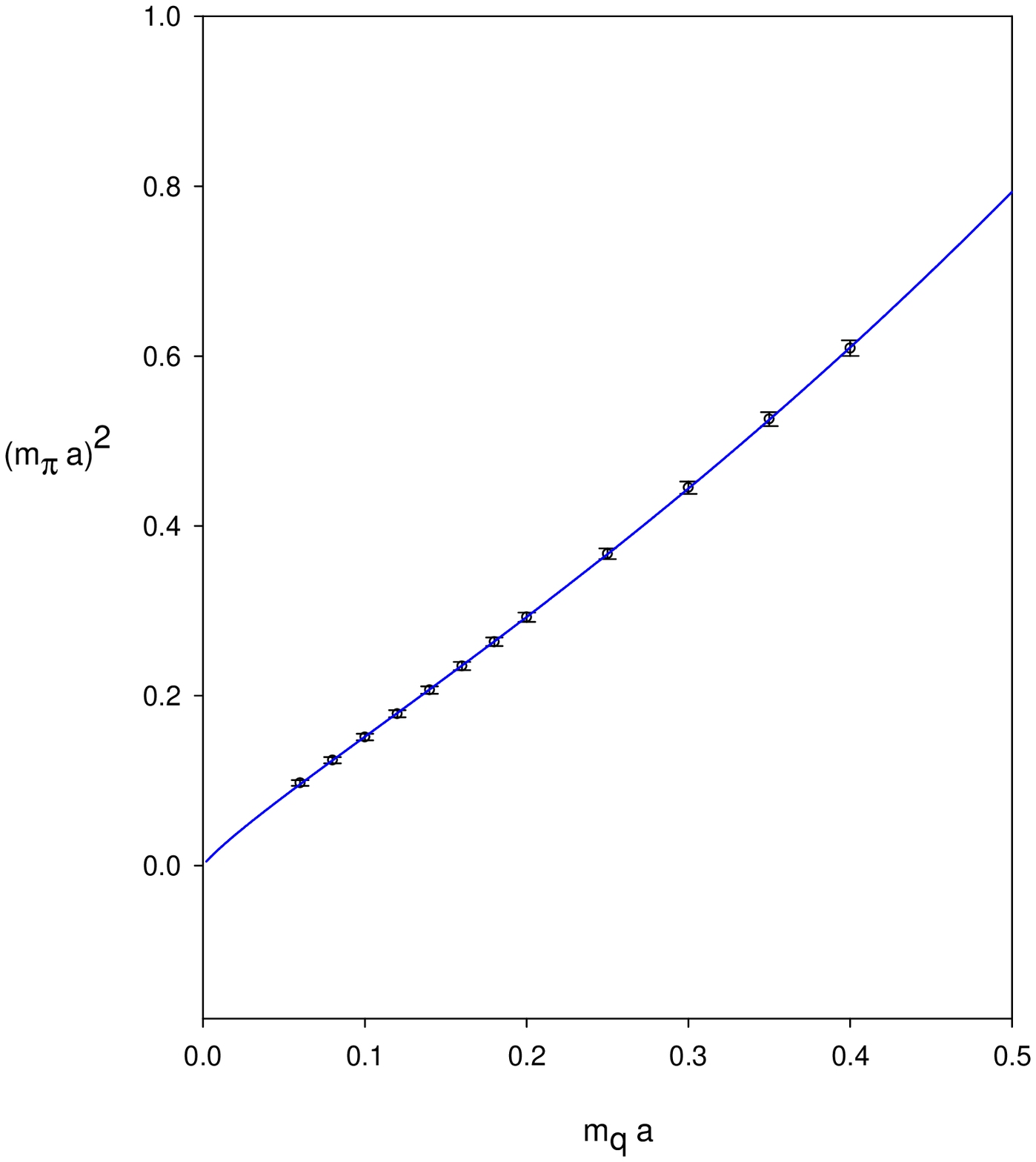} {
The pion mass square
$ ( m_{\pi} a  )^2  $ versus the bare quark mass $ m_q a $.
The solid line is the fit of Eq. (\ref{eq:mpi2}). }

\psfigure 5.0in -0.2in {fig:fpi} {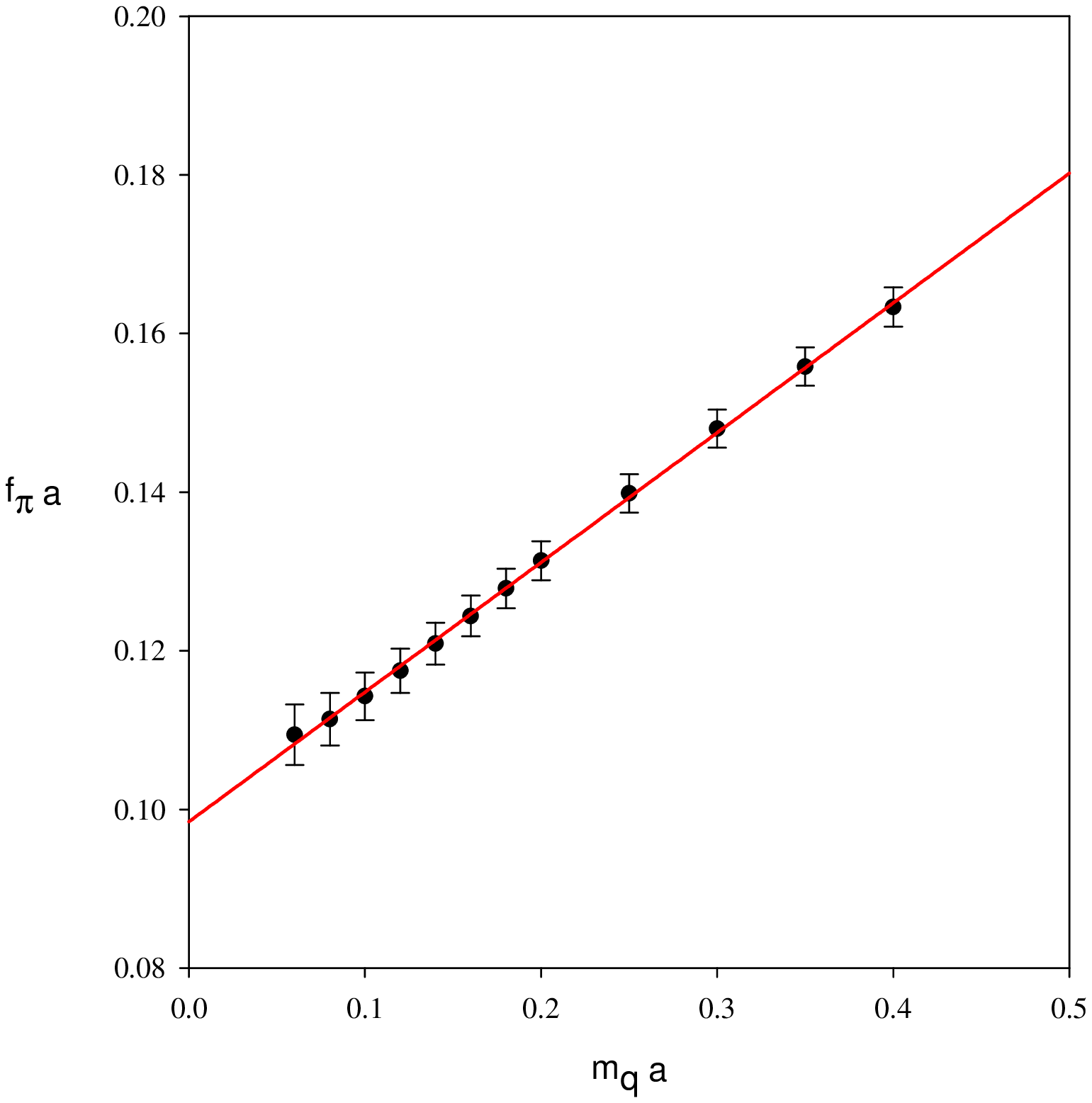} {
The pion decay constant $ f_{\pi} a $ versus the bare quark mass
$ m_q a $. The solid line is the linear fit.}

\psfigure 5.0in -0.2in {fig:mpi2omq} {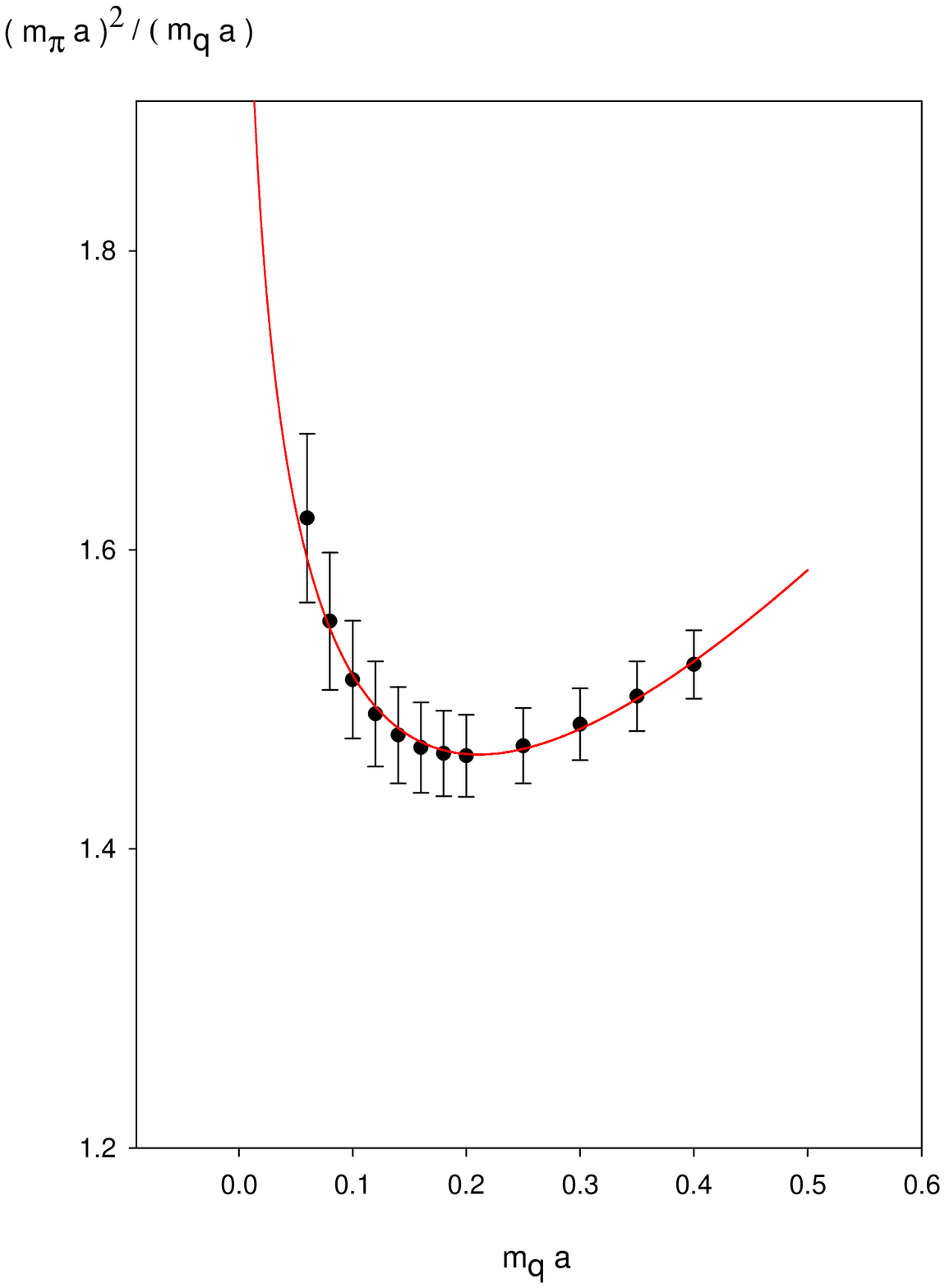} {
$ ( m_{\pi} a )^2 / ( m_q a ) $ versus the bare quark mass $ m_q a $.
The solid line is the fit of Eq. (\ref{eq:mpi2}) divided by $ m_q a $.}

\psfigure 5.0in -0.2in {fig:chilog} {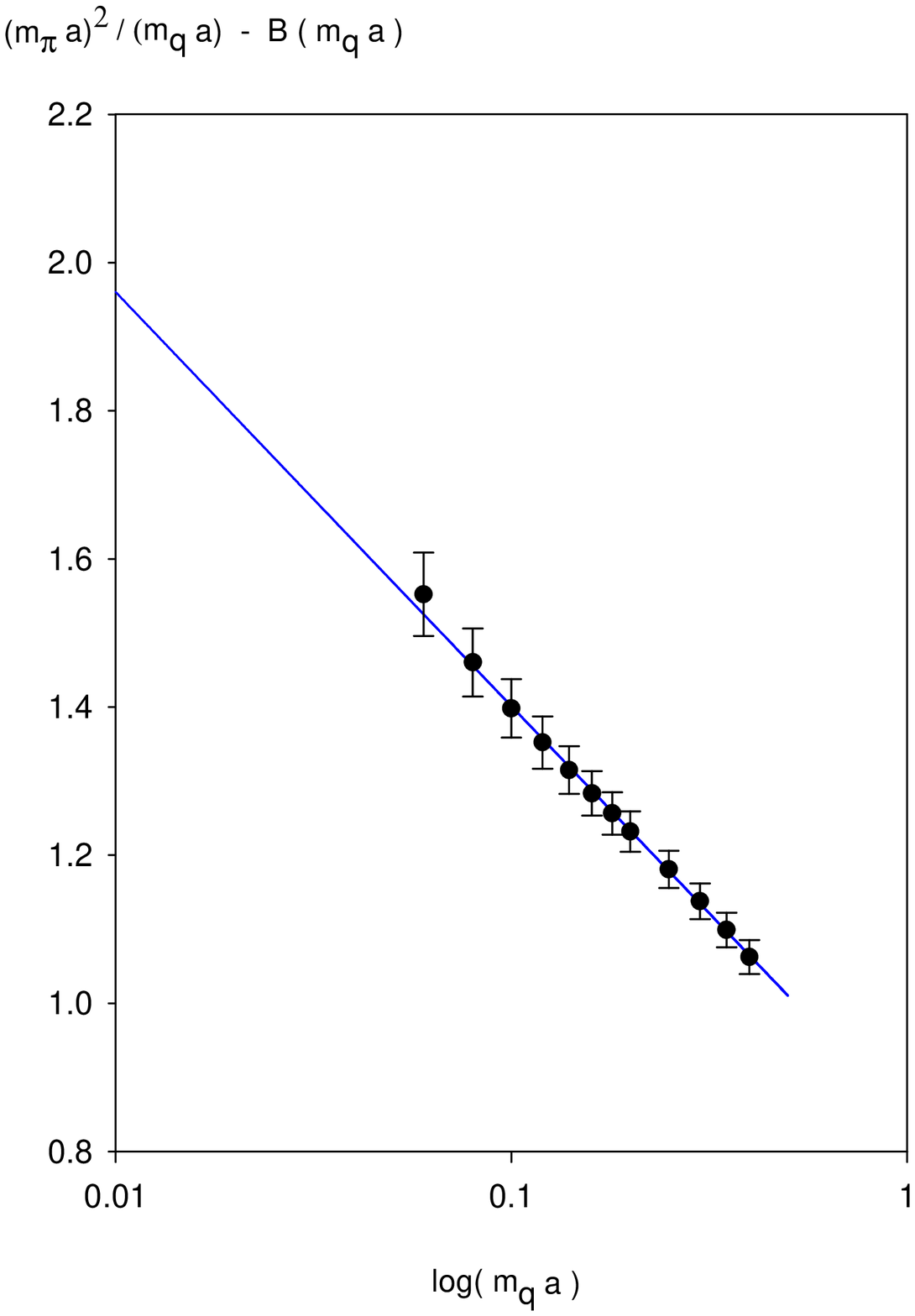} {
The extraction of quenched chiral logarithm by
plotting $ ( m_{\pi} a )^2 / (m_q a )-B( m_q a ) $ versus $ \log( m_q a ) $.
From the slope of the fitted straight line,
the coefficient of quenched chiral logarithm can also be determined to be
$ \delta = 0.2034 \pm 0.0140 $. }

In passing, we note that a finite lattice must impose a lower bound for
the pion mass, as well as the corresponding one for the bare quark mass.
If one decreases the bare quark mass beyond its lower bound, then the
resulting pion mass would be flattening ( or increasing ) rather than
decreasing. This is essentially due to the finite volume effects of the
zero modes of the overlap Dirac operator in the pion propagator, which
is proportional to $ |n_{+} - n_{-}|/( m_q^2 a^2 N_s ) $ as $ m_q a \to 0 $.
Thus, only in the infinite volume limit ( $ N_s \to \infty $ ),
one can obtain zero pion mass with zero bare quark mass.
Nevertheless, for a finite lattice, the finite volume effects of
the zero modes can be suppressed if the bare quark mass
$ m_q a \gg \sqrt{| n_{+} - n_{-} | / N_s } $.
For a lattice of size $ 8^3 \times 24 $ at $ \beta = 5.8 $,
we find that $ m_q a \ge 0.06 $ is sufficient to suppress the
finite volume effects of the zero modes. This can be verified by
comparing the results (\ref{eq:delta_pi})-(\ref{eq:B}) to those
obtained in larger volumes, e.g., $ 12^3 \times 24 $ at $ \beta = 5.8 $
( which is about $ 3.4 $ times of the present volume ). Another plausible
check is to see whether the $ \delta $ extracted from the pion mass agrees
with that obtained from the index susceptibility, Eq.(4), since
the index susceptibility and the $ f_\pi a $ ( see Fig. 2 ) both
presumably suffer the least from the finite volume effects.
The latter check is performed in Section 3, and the agreement
between these two $ \delta's $ suggests that the finite volume
effects may be under control.

\section{Index susceptibility and the $\eta'$ mass }

In this section, we derive a relationship between
the index susceptibility of any Ginsparg-Wilson lattice Dirac operator
and the $ \eta' $ mass as depicted in quenched chiral perturbation theory.
Then we measure the index of overlap Dirac operator for those
100 gauge configurations used for computing quark propagators in Section 2,
and obtain the index susceptibility, the $ \eta' $ mass, and
an estimate of $ \delta = 0.197 \pm 0.027 $,
which is in good agreement with the $ \delta $ determined
from the pion mass in Section 2.

\subsection{A formula due to exact chiral symmetry}

The propagator of ( flavor-singlet ) $ \eta' $ can be written as
\bea
\label{eq:eta'}
M_{\eta'} (x,y) &=&
\tr \{\gamma_5 (D_c + m_q)^{-1} (y,x) \gamma_5 (D_c + m_q)^{-1} (x,y) \} \nn
& & - \tr \{\gamma_5 (D_c + m_q)^{-1} (y, y) \}
      \tr \{\gamma_5 (D_c + m_q)^{-1} (x, x) \}
\eea
where the second term on r.h.s. is the commonly called hairpin diagram.
Note that the first term is just the usual pion propagator (\ref{eq:pion}),
while the second term can be regarded as pion propagator with the
$ \eta' $ mass insertion due to the nontrivial interactions of gluons
between two seemingly disconnected quark propagators.

In the quenched approximation, the expectation value of the
hairpin diagram in momentum space can be written as
\bea
\label{eq:hairpin}
& & \left< \frac{1}{V}
    \sum_{x,y} e^{i p \cdot (x-y) }
           \ \tr \{\gamma_5 (D_c + m_q)^{-1} (y, y) \}
             \tr \{\gamma_5 (D_c + m_q)^{-1} (x, x) \} \right> \nn
&=& \sqrt{Z}\frac{1}{p^2+m_\pi^2} m_{\eta'}^2 \frac{1}{p^2+m_\pi^2} \sqrt{Z}
\eea
where the brackets $ \langle \cdots \rangle $ denote quenched average
over gauge configurations with weight $ e^{-{\cal A}_g } $
( $ {\cal A}_g $ : pure gauge action ), and
\BAN
\label{eq:Z}
\sqrt{Z} = \frac{m_\pi^2 f_\pi}{ 2 m_q }
\EAN
as defined in (\ref{eq:Gt_fit}) and (\ref{eq:fpi}).
Equation (\ref{eq:hairpin}) can be regarded as the definition of the
$ \eta' $ mass in quenched chiral perturbation theory.

At zero momentum $ p = 0 $, (\ref{eq:hairpin}) becomes
\bea
\label{eq:hairpin_0}
\left< \frac{1}{V}
    \sum_{x,y}
         \ \tr \{\gamma_5 (D_c + m_q)^{-1} (y, y) \}
           \tr \{\gamma_5 (D_c + m_q)^{-1} (x, x) \} \right>
= \frac{ f_{\pi}^2 m_{\eta'}^2 }{ 4 m_q^2}
\eea

Now the expression inside the brackets $ \langle \cdots \rangle $
can be evaluated exactly in terms of the index of $ D_c $ ( $ D $ ).
Explicitly,
\bea
\label{eq:trace}
& & \sum_{x} \tr \{ \gamma_5 (D_c + m_q)^{-1} (x,x) \}
 = \Tr \{ \gamma_5 (D_c + m_q)^{-1} \} \nn
&=& ( 1 - r m_q a )^{-1}
    \sum_{\alpha} \phi_{\alpha}^{\dagger} \gamma_5 \phi_{\alpha}
    \{ [ m_q + (1 - r m_q a ) \lambda_{\alpha} ]^{-1} - r a \}
\eea
where (\ref{eq:Dm}) and (\ref{eq:Dcm}) have been used, and
the trace $ \mbox{Tr} $ is evaluated in terms of the eigenvalues
$ \{ \lambda_{\alpha} \} $ and eigenvectors $ \{ \phi_{\alpha} \} $
of the Ginsparg-Wilson lattice Dirac operator $ D $
(\ref{eq:gen_sol}) with $ D_c $ satisfying $ (i D_c )^{\dagger} = i D_c $.
Recalling the properties of the eigensystem of $ D $ \cite{Chiu:1998bh},
\BAN
  \phi_{\alpha}^{\dagger} \gamma_5 \phi_{\alpha}
= \left\{  \begin{array}{ll}
     0      &  \mbox{ for $ \lambda_{\alpha} \ne \lambda_{\alpha}^{*} $ } \\
   \pm 1    &  \mbox{ for $ \lambda_{\alpha}  =  \lambda_{\alpha}^{*} $ } \\
             \end{array}  \right.
\EAN
one immediately sees that only the real eigenmodes
$ \lambda = 0 $ and $ (ra)^{-1} $ can
contribute to (\ref{eq:trace}). Since the factor
$ \{ [ m_q + (1 - r m_q a ) \lambda_{\alpha} ]^{-1} - r a \} $ vanishes
for $ \lambda_{\alpha} = (ra)^{-1} $, thus only the
zero modes contribute to (\ref{eq:trace}), i.e.,
\bea
\label{eq:g5qloop}
 \sum_{x} \tr \{ \gamma_5 (D_c + m_q)^{-1} (x,x) \}
=\frac{n_{+} - n_{-}}{m_q} \ .
\eea
Substituting (\ref{eq:g5qloop}) into l.h.s. of (\ref{eq:hairpin_0}),
one obtains
\bea
\label{eq:wv_lat}
   N_f \frac{ \langle (n_{+} - n_{-})^2 \rangle }{V}
=  \frac{ f_{\pi}^2  m_{\eta'}^2 }{4}  \ ,
\eea
where the factor $ N_f $ denotes the number of light quark flavors,
which accounts for the number of hairpin diagrams contributing
to the $ \eta' $ mass.
Since (\ref{eq:wv_lat}) is an exact result following from the definition of
$ m_{\eta'} $ in (\ref{eq:hairpin}), it offers the best way to obtain the
$ \eta' $ mass in quenched chiral perturbation theory.
One just measures the index ( susceptibility ) of the Ginsparg-Wilson
lattice Dirac operator, without computing the hairpin diagram at all.
This is one of the advantages of preserving exact chiral symmetry
on the lattice. Note that (\ref{eq:wv_lat}) is independent of the
bare quark mass $ m_q $, thus the $ \eta' $ can remain massive even
in the limit $ m_q \to 0 $. This is the crucial feature of the
$ \eta' $ to distinguish itself from the octet of approximate Goldstone
bosons, the $ \pi, K $ and $ \eta $.

Recently, (\ref{eq:wv_lat}) is derived \cite{Giusti:2001xh} differently
starting from the anomalous flavor-singlet axial Ward identity,
which is formally equivalent to replacing the continuum topological
charge density $ \rho $ in the Witten-Veneziano formula
\cite{Witten:1979vv,Veneziano:1979ec}
\bea
\label{eq:wv_cont}
  m_{\eta'}^2
= \frac{4 N_f}{ f_{\pi}^2 } \int d^4 x \langle \rho(x) \rho(0) \rangle \ ,
\hspace{4mm}
\rho =  \frac{1}{32\pi^2} \epsilon_{\mu\nu\lambda\sigma}
            \tr(F_{\mu\nu} F_{\lambda\sigma} ) \ ,
\eea
with the axial anomaly $ q = \tr[ \gamma_5( 1 - ra D ) ] $
of the overlap Dirac operator, and the integration $ \int d^4 x $
with the summation over all sites $ \sum_x $. That is, the integral
in (\ref{eq:wv_cont}) is formally transcribed as
\bea
\label{eq:wvv}
 \int d^4 x \langle \rho(x) \rho(0) \rangle \Longleftrightarrow
 a^{-4} \sum_x \langle q(x) q(0) \rangle \ .
\eea
Since (\ref{eq:wv_cont}) is derived in the large $ N_c $ limit
( which is valid to lowest order in $ 1/N_c $ ),
a relevant question is what value of $ N_c $ in Eq. (\ref{eq:wv_cont})
is supposed to be valid. Further, even though the r.h.s. of (\ref{eq:wvv})
is well-defined on a finite lattice, some subtleties \cite{Seiler:2001je}
could emerge as the 2-point function
$ \langle q(x) q(0) \rangle $ ( $ x \ne 0 $ ) becomes non-positive in
the continuum limit, since $ \rho(x) $ is odd under time reflections.
Thus, a real $ m_{\eta'} $ would require a
divergent contact term $ \langle \rho^2(0) \rangle $.
How to fix this contact term is the basic problem of Witten-Veneziano
formula (\ref{eq:wv_cont}) in the continuum. On the other hand,
for a finite lattice, one can use the well-known relation
$ \sum_x q(x) = n_{+} - n_{-} $ to turn the r.h.s. of (\ref{eq:wvv})
into the index susceptibility,
\bea
\label{eq:qqs}
 a^{-4} \sum_x \langle q(x) q(0) \rangle
= \frac{1}{V} \sum_x \sum_y \langle q(x) q(y) \rangle
= \frac{1}{V} \langle ( n_{+} - n_{-} )^2 \rangle \ ,
\eea
which is well-defined in the limit $ V \to \infty $ and
$ a \to 0 $. Nevertheless, it is still interesting to see
whether $ \langle q(x) q(0) \rangle $ ( $ x \ne 0 $ ) would
become negative in the continuum limit, together with a divergent
contact term.

It should be emphasized that our derivation of (\ref{eq:wv_lat})
is quite different from that of Ref. \cite{Giusti:2001xh}.
We start from Eq. (\ref{eq:hairpin}) which
serves as the definition of the $ \eta' $ mass
in quenched chiral perturbation theory. Then the subsequent steps
leading to (\ref{eq:wv_lat}) are exact due to chiral symmetry,
without any approximations or assumptions.
This is in contrast to the approach used in Ref. \cite{Giusti:2001xh},
in which the anomalous flavor-singlet axial Ward identity is used,
in the limit $ u \equiv N_f / N_c \to 0 $. The subtleties
arising from $ \langle q(x) q(0) \rangle $ in the continuum limit
cannot be resolved unless one transcribes the expression in terms of the
index susceptibility of a Ginsparg-Wilson lattice Dirac operator.
Finally we note that a derivation of (\ref{eq:wv_lat}) similar to our
approach is also presented in Ref. \cite{DeGrand:2002gm}
for the overlap Dirac operator.

\subsection{Measurement of the index of the overlap}

The index of overlap Dirac operator (\ref{eq:overlap}) is
\bea
\label{eq:ind_D}
\mbox{index}( D ) = \sum_{x} \tr[ \gamma_5 ( 1 - a r D(x,x)) ]
                  = \frac{1}{2} ( h_{-} - h_{+} )
\eea
where $ h_{+} ( h_{-} ) $ is the number of positive ( negative )
eigenvalues of the hermitian Wilson-Dirac operator $ H_w $ (\ref{eq:Hw}).
However, one does not need to obtain all eigenvalues of $ H_w $
in order to know how many of them are positive or negative.
The idea is simple. Since $ H_w $ has equal number of positive and
negative eigenvalues for $ m_0 \le 0 $, then one can just focus at those
low-lying ( near zero ) eigenmodes of $ H_w $, and see whether any of
them crosses zero from positive to negative, or vice versa, when $ m_0 $
is scanned from $ 0 $ up to the value ( e.g., $ 1.30 $ in this paper )
used in the definition of $ D $. From the net number of crossings,
one can obtain the index of $ D $ (\ref{eq:ind_D}). This is the spectral
flow method used in Refs. \cite{Narayanan:1995gw,Edwards:1998sh}
to obtain the index of overlap Dirac operator.

In Fig. \ref{fig:050}, we plot the spectral flow of eight lowest-lying
( near zero ) eigenvalues of $ H_w( m_0) $ in the interval
$ 0.9 \le m_0 \le 1.3 $, for one of the gauge configurations.
Evidently, the flows are not as smooth as one may have expected.
In this case, the net crossings is -6 ( from negative to positive ),
so the index of the overlap Dirac operator is -6.
In some cases ( see, for example, Fig \ref{fig:008} ), there are some
intriguing eigenvalues lying very close to zero, thus for a coarse
scan in $ m_0 $, it may not be so easy to determine whether they actually
cross zero or not. These ambiguities can only be resolved by
tracing them closely at a finer resolution in $ m_0 $.

\psfigure 5.0in -0.2in {fig:050} {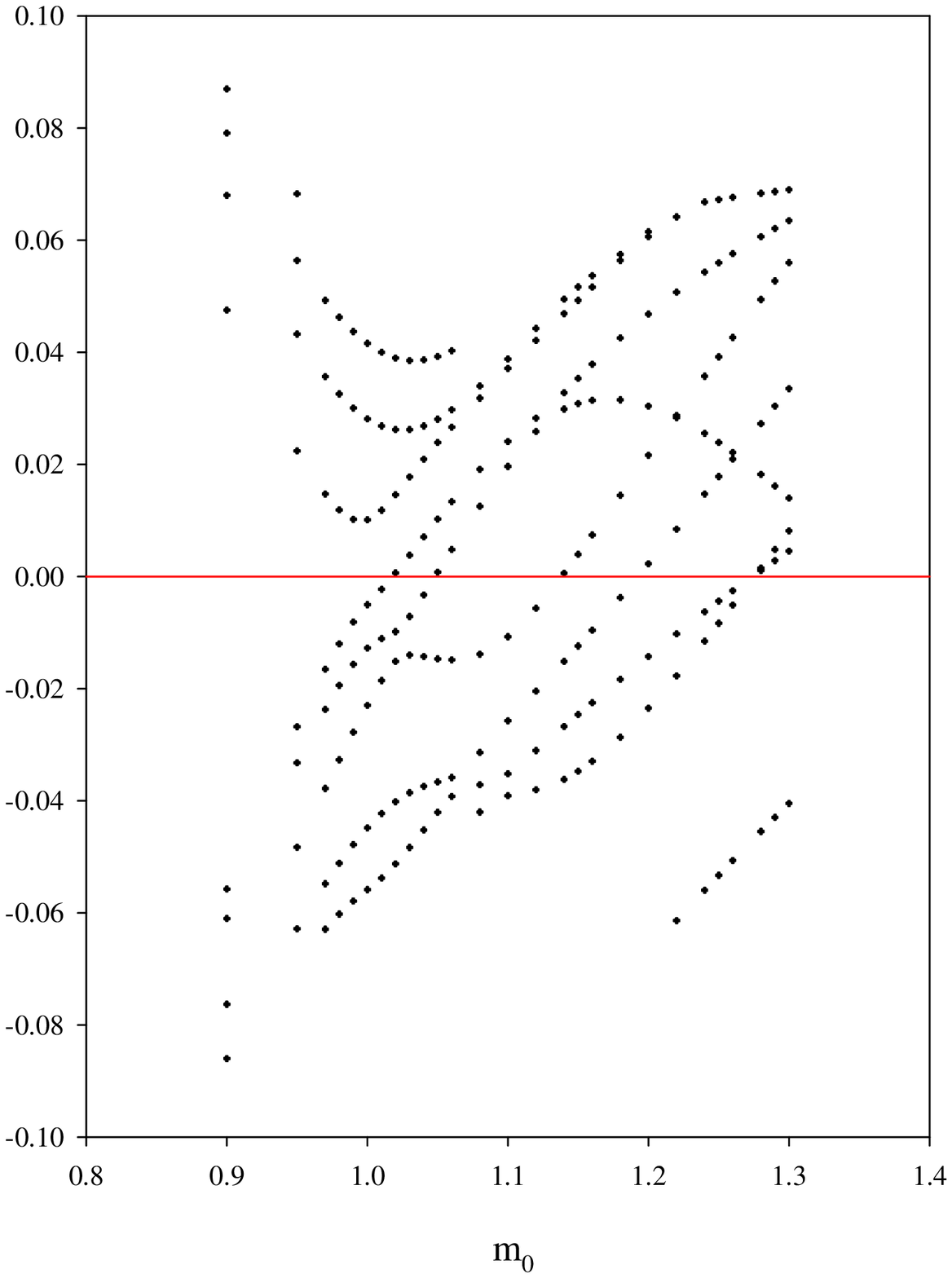} {
The spectral flow of 8 lowest-lying eigenvalues of $ H_w $ for the
50th gauge configuration. There are 6 crossings
from negative to positive, so the index is equal to -6. }

\psfigure 5.0in -0.2in {fig:008} {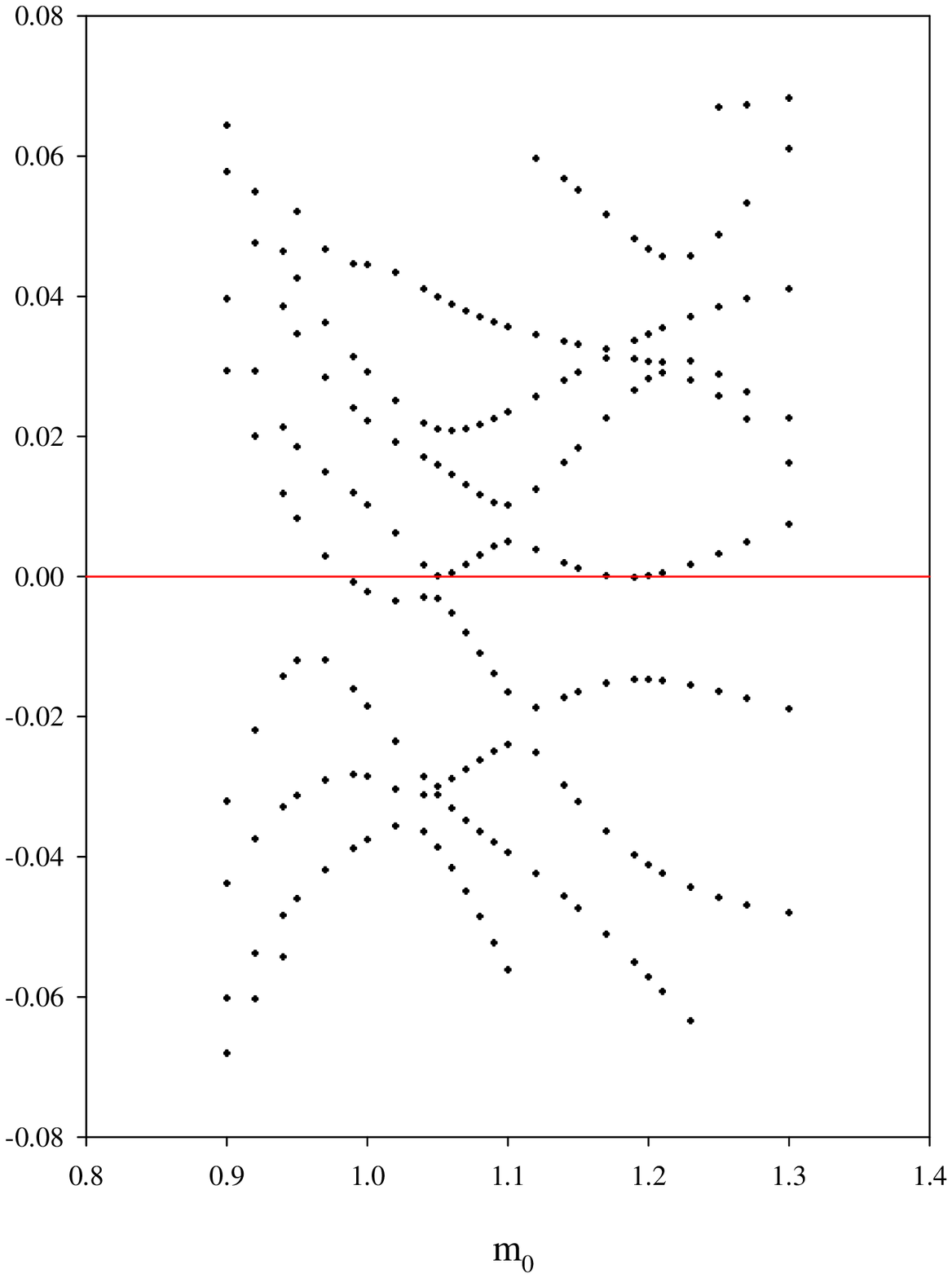} {
The spectral flow of 8 lowest-lying eigenvalues of $ H_w $ for
the 8th gauge configuration. There is only one crossing
from positive to negative around $ m_0 = 0.98 $, so the index is equal to
+1.}

The distribution of the indices of $ D $ for these
100 gauge configurations is listed in Table \ref{table:index}.
So the index susceptibility is
\bea
\label{eq:index_s}
a^4 \chi =
\frac{ \left< ( n_{+} - n_{-} )^2 \right> }{N_s} = 3.67(50) \times 10^{-4}
\eea
where $ N_s $ is the total number of sites. Now if we subtract
the mean value $  \langle n_{+} - n_{-} \rangle = -0.19 $ from all indices,
then the index susceptibility becomes
\bea
\label{eq:index_ss}
a^4 \chi =
\frac{ \left< ( n_{+} - n_{-} )^2 \right> -
       \left<  n_{+} - n_{-} \right>^2 }{N_s} = 3.64(50) \times 10^{-4}
\eea

{\footnotesize
\begin{table}
\begin{center}
\begin{tabular}{|c|c|}
\hline
$ I = n_{+} - n_{-} $   &   number of configurations    \\
\hline
\hline
     4    &       2   \\
\hline
     3    &      11   \\
\hline
     2    &       9   \\
\hline
     1    &      14   \\
\hline
     0    &      24   \\
\hline
    -1    &      15   \\
\hline
    -2    &      11   \\
\hline
    -3    &       6   \\
\hline
    -4    &       6   \\
\hline
    -5    &       1   \\
\hline
    -6    &       1   \\
\hline
\hline
\end{tabular}
\end{center}
\caption{
The distribution of the indices of overlap Dirac operator
for 100 gauge configurations at $ \beta = 5.80 $ on the
$ 8^3 \times 24 $ lattice.
Here $ \langle I \rangle =-0.19 $ and $ \langle I^2 \rangle =4.51 $.}
\label{table:index}
\end{table}
}

Now substitute the index susceptibility (\ref{eq:index_ss}),
the lattice spacing (\ref{eq:a}), $ f_\pi = 132 $ MeV,
and $ N_f = 3 $, into the exact relation (\ref{eq:wv_lat}), we
obtain the $ \eta' $ mass
\bea
m_{\eta'} =  ( 901 \pm 64 ) \mbox{MeV} \ ,
\eea
which agrees with the theoretical estimate
\bea
\label{eq:eta_expt}
\sqrt{ {\bf m}_{\eta'}^2 + m_{\eta}^2 - 2 m_{K}^2 } = 853  \mbox{ MeV } \ ,
\eea
with experimental values of meson masses\footnote{Here
we distinguish the physical $ \eta' $ mass ( $ {\bf m}_{\eta'} $ ),
from the $ \eta' $ mass ( $ m_{\eta'} $ ) in queneched chiral perturbation
theory. } :
$ {\bf m_{\eta'} } = 958 $ MeV, $ m_{\eta} = 547 $ MeV,
and $ m_{K} = 495 $ MeV.

Next we substitute (\ref{eq:index_ss}) and
$ f_{\pi} a = 0.0984(3) $ into (\ref{eq:delta_s}), and get
\bea
\label{eq:delta_is}
\delta = 0.197 \pm 0.027 \ ,
\eea
which is in good agreement with the value (\ref{eq:delta_pi})
determined from the pion mass, as well as with the theoretical estimate
$ \delta \simeq 0.2 $.

\section{Concluding remarks}

In this paper, we have examined the quenched chiral logarithm
in lattice QCD with overlap Dirac quark. The coefficient
of quenched chiral logarithm ( $ \delta = 0.203 \pm 0.014 $ )
extracted from the pion mass agrees very well with that
( $ \delta = 0.197 \pm 0.027 $ ) obtained from the index susceptibility
of overlap Dirac operator. Further, they are in good agreement with the
theoretical estimate $ \delta \simeq 0.2 $ in quenched chiral perturbation
theory. This provides strong evidences that lattice QCD with overlap
Dirac quark realizes quenched QCD chiral dynamics, as depicted by
quenched chiral perturbation theory.

Our results on pion mass and decay constant rely very much
on the viability of computing quark propagators to a high
accuracy. We find that the accuracy in the implementation of
the inverse square root of $ H_w^2 $ is the most crucial step
in this two-level conjugate gradient paradigm.
The Zolotarev optimal rational approximation together
with projection of high and low-lying eigenmodes enables us
to control the error of $ \mbox{sign}(H_w) Y $ always
less than $ 10^{-11} $, at each iteration of the outer CG loop.
The details of our implementation are described in Section 2.

Equation (\ref{eq:wv_lat}) is an exact result following from the
definition of the $ \eta'$ mass in (\ref{eq:hairpin}).
It provides the best way to obtain the
$ \eta' $ mass in quenched chiral perturbation theory, as well as
to determine the coefficient of quenched chiral logarithm in terms of
the index susceptibility of any topologically-proper
Ginsparg-Wilson lattice Dirac operator.
Equation (\ref{eq:delta_s}) may even suggest that $ \delta $
is scale invariant for a range of lattice spacings including
the continuum limit $ a \to 0 $. This also explains why the
$ \delta $ value we obtained at $ a = 0.147 $ fm is so close to the
theoretical estimate in the continuum.

At this point, it may be interesting to recall an example of lattice Dirac
operator with exact chiral symmetry \cite{Chiu:2001bg}. Unlike the overlap
Dirac operator, it does not have topological zero modes for any nontrivial
gauge backgrounds. However, it reproduces correct axial anomaly in the
continuum limit \cite{Chiu:2001ja}, at least for the trivial gauge sector.
Since its index is zero, its index susceptibility must be zero,
and it follows that the $ \eta' $ mass is zero, and the quenched chiral
logarithm is absent in the pion mass as well as other related quantities.
Nevertheless, it does not necessarily imply that this lattice
Dirac operator could not be realized in nature. It seems that the real
testing ground for this lattice Dirac operator is lattice QCD with
dynamical quarks, in which all quenched pathologies are absent.

In passing, we briefly outline the essential features of our computing
system. The platform is a home-made Linux PC cluster with 18 nodes, built
with off-the-shelf components. Each node consists of one Pentium 4
processor ( 1.6 Ghz ) with one Gbyte of Rambus, one $ 40 $ Gbyte
hard disk, and a network card. The computational intensive parts
( matrix-vector operations ) of our program are written in SSE2
codes \cite{Luscher:2001tx}, which can double the speed of
our earlier pure Fortran codes compiled by Intel Fortran Compiler 5.0
with maximum optimizations including the SSE2 option.
The performance of our system is estimated to be around 20 Gflops.




This work was supported in part by the National Science Council,
ROC, under the grant number NSC90-2112-M002-021,
and also in part by NCTS.

\bigskip
\bigskip



\begin{thebibliography}{15}

\bibitem{Sharpe:1992ft}
S.~R.~Sharpe,
Phys.\ Rev.\ D {\bf 46}, 3146 (1992)


\bibitem{Bernard:1992mk}
C.~W.~Bernard and M.~F.~Golterman,
Phys.\ Rev.\ D {\bf 46}, 853 (1992)


\bibitem{Aoki:1999yr}
S.~Aoki {\it et al.}  [CP-PACS Collaboration],
Phys.\ Rev.\ Lett.\  {\bf 84}, 238 (2000)

\bibitem{Bardeen:2000cz}
W.~Bardeen, A.~Duncan, E.~Eichten and H.~Thacker,
Phys.\ Rev.\ D {\bf 62}, 114505 (2000)

\bibitem{Bernard:2001av}
C.~W.~Bernard {\it et al.},
Phys.\ Rev.\ D {\bf 64}, 054506 (2001)


\bibitem{Neuberger:1998fp}
H.~Neuberger,
Phys.\ Lett.\ B {\bf 417}, 141 (1998);

\bibitem{Narayanan:1995gw}
R.~Narayanan and H.~Neuberger,
Nucl.\ Phys.\ B {\bf 443}, 305 (1995)


\bibitem{Dong:2001fm}
S.~J.~Dong, T.~Draper, I.~Horvath, F.~X.~Lee, K.~F.~Liu and J.~B.~Zhang,
Phys.\ Rev.\ D {\bf 65}, 054507 (2002)


\bibitem{Ginsparg:1982bj}
P.~H.~Ginsparg and K.~G.~Wilson,
Phys.\ Rev.\ D {\bf 25}, 2649 (1982).

\bibitem{Chiu:1998gp}
T.~W.~Chiu and S.~V.~Zenkin,
Phys.\ Rev.\ D {\bf 59}, 074501 (1999)

\bibitem{Chiu:1998eu}
T.~W.~Chiu,
Phys.\ Rev.\ D {\bf 60}, 034503 (1999)


\bibitem{Chiu:1998ce}
T.~W.~Chiu,
Phys.\ Rev.\ D {\bf 60}, 114510 (1999)

\bibitem{Neuberger:1998my}
H.~Neuberger,
Phys.\ Rev.\ Lett.\  {\bf 81}, 4060 (1998)

\bibitem{Frommer:1995ik}
A.~Frommer, B.~Nockel, S.~Gusken, T.~Lippert and K.~Schilling,
Int.\ J.\ Mod.\ Phys.\ C {\bf 6}, 627 (1995)


\bibitem{Zol:1877}
E.~I.~Zolotarev,
``Application of elliptic functions to the questions of functions
deviating least and most from zero'',
Zap. Imp. Akad. Nauk. St. Petersburg, 30 (1877), no. 5; reprinted in his
Collected works, Vol. 2, Izdat, Akad. Nauk SSSR, Moscow, 1932, p. 1-59.


\bibitem{vandenEshof:2001hp}
J.~van den Eshof, A.~Frommer, T.~Lippert, K.~Schilling and H.~A.~van der Vorst,
Nucl.\ Phys.\ Proc.\ Suppl.\  {\bf 106}, 1070 (2002)

\bibitem{Chiu:1998bh}
T.W.~Chiu,
Phys.\ Rev.\ D {\bf 58}, 074511 (1998)


\bibitem{Witten:1979vv}
E.~Witten,
Nucl.\ Phys.\ B {\bf 156}, 269 (1979).

\bibitem{Veneziano:1979ec}
G.~Veneziano,
Nucl.\ Phys.\ B {\bf 159}, 213 (1979).


\bibitem{Giusti:2001xh}
L.~Giusti, G.~C.~Rossi, M.~Testa and G.~Veneziano,
Nucl.\ Phys.\ B {\bf 628}, 234 (2002)


\bibitem{Seiler:2001je}
E.~Seiler,
Phys.\ Lett.\ B {\bf 525}, 355 (2002);
E.~Seiler and I.~O.~Stamatescu,
MPI-PAE/PTh 10/87.


\bibitem{DeGrand:2002gm}
T.~DeGrand and U.~M.~Heller  [MILC collaboration],
hep-lat/0202001.


\bibitem{Creutz:zw}
M.~Creutz,
Phys.\ Rev.\ D {\bf 21} (1980) 2308.

\bibitem{Cabibbo:zn}
N.~Cabibbo and E.~Marinari,
Phys.\ Lett.\ B {\bf 119} (1982) 387.

\bibitem{Edwards:1998sh}
R.~G.~Edwards, U.~M.~Heller and R.~Narayanan,
Nucl.\ Phys.\ B {\bf 535}, 403 (1998)

\bibitem{Chiu:2001bg}
T.~W.~Chiu,
Phys.\ Lett.\ B {\bf 521}, 429 (2001)




\bibitem{Chiu:2001ja}
T.~W.~Chiu and T.~H.~Hsieh,
Phys.\ Rev.\ D {\bf 65}, 054508 (2002)


\bibitem{Luscher:2001tx}
M.~Luscher,
Nucl.\ Phys.\ Proc.\ Suppl.\  {\bf 106}, 21 (2002)


\end{thebibliography}
\end{document}